# Dynamics for droplet-based electricity generators

Xiang Wang, Sunmiao Fang, Jin Tan, Tao Hu, Weicun Chu, Jun Yin, Jianxin Zhou, Wanlin Guo*

*Key Laboratory for Intelligent Nano Materials and Devices of the Ministry of Education, State Key Laboratory of Mechanics and Control of Mechanical Structures, Institute of Nanoscience, Nanjing University of Aeronautics and Astronautics, Nanjing 210016, China*

***Corresponding author.** E-mail: wlguo@nuaa.edu.cn

## ABSTRACT

**The finding of droplet-based electricity generator[1] (DEG), based on the moving boundary of electrical double layer, has triggered great research enthusiasm[2-19], and a breakthrough in instantaneous electric power density was achieved recently[19]. However, the dynamic mechanism for such droplet-based electricity generators remains elusive, impeding optimization of the DEG for practical applications. Through comprehensive experiments, we developed a dynamic model of surface charge density that can explain the underlying mechanism for the DEGs. The spreading droplet in touch with the top electrode can be equivalently regarded as an additional part of the top plate of the DEG capacitor, and the change of droplet area causes the change of surface charge density of the plates, driving electrons to migrate between the two plates. The insight of the dynamic mechanism paves a way for optimal design and practical applications of DEGs.**

**Keywords:** droplet-based electricity generator (DEG), moving boundary, dynamic model, surface charge density

## Introduction

Harvesting the tremendous energy contained in water directly through its interaction with functional materials has been drawing increasing interesting and research enthusiasm[1-19], leading to the emerging hydrovoltaic technology[10], converting various forms of energy from water to electric power. Among these techniques, the droplet-based electricity generator (DEG) has attracted significant attention, which is capable to generate higher instantaneous electric power density using simpler design. But the underlying mechanism of some EDGs remains elusive. Recently, a mathematical RC circuit model was proposed to describe the electrical signal of a particular EDG quantitatively[20], however, it is not the inverse of electrowetting conceptually and is lack of a dynamic understanding. To further develop DEGs and transform them to practical applications, a clear mechanism is vitally required.

Here we perform comprehensive experimental investigation using a DEG similar to that reported previously[19], which can generate electricity repeatedly by dropping a series of droplets. Based on observation of the dynamics of the droplet by a high-speed camera and measurement of its induced electric signals, we develop a dynamic model of surface charge density for the working mechanism of this DEG. The dynamic model paves the way for the further optimization of DEGs.

## Results

The device structure is illustrated in Fig.1a, where a 15 mm×2 mm piece of copper electrode lies on the top surface of a PTFE film with a dimension of 25 mm×75 mm×0.03 mm, whose back side was coated with gold electrode. The PTFE film was attached to a class slide for support. As shown in Fig.1b, one tap droplet (ion concentration around 3mM) generated a sharp positive voltage pulse up to 143.8 V and a followed negative voltage signal up to -17.1 V (The droplet was released at a height of 17 cm, while the device was placed with a tilt angle of 35°). It's worth noting that there was a negative pulse following the positive pulse, which means a back flow of electrons. Before the droplet contacts the electrode, no voltage is generated. At $T_{touch}$, when the droplet touches the top electrode, a sharp voltage peak is suddenly induced. Following that, the voltage changes while the droplet area changes dynamically (Fig. 1c). The output electric power is high enough to instantaneously light up a light-emitting diodes (LED), as shown in Supplementary video 1.

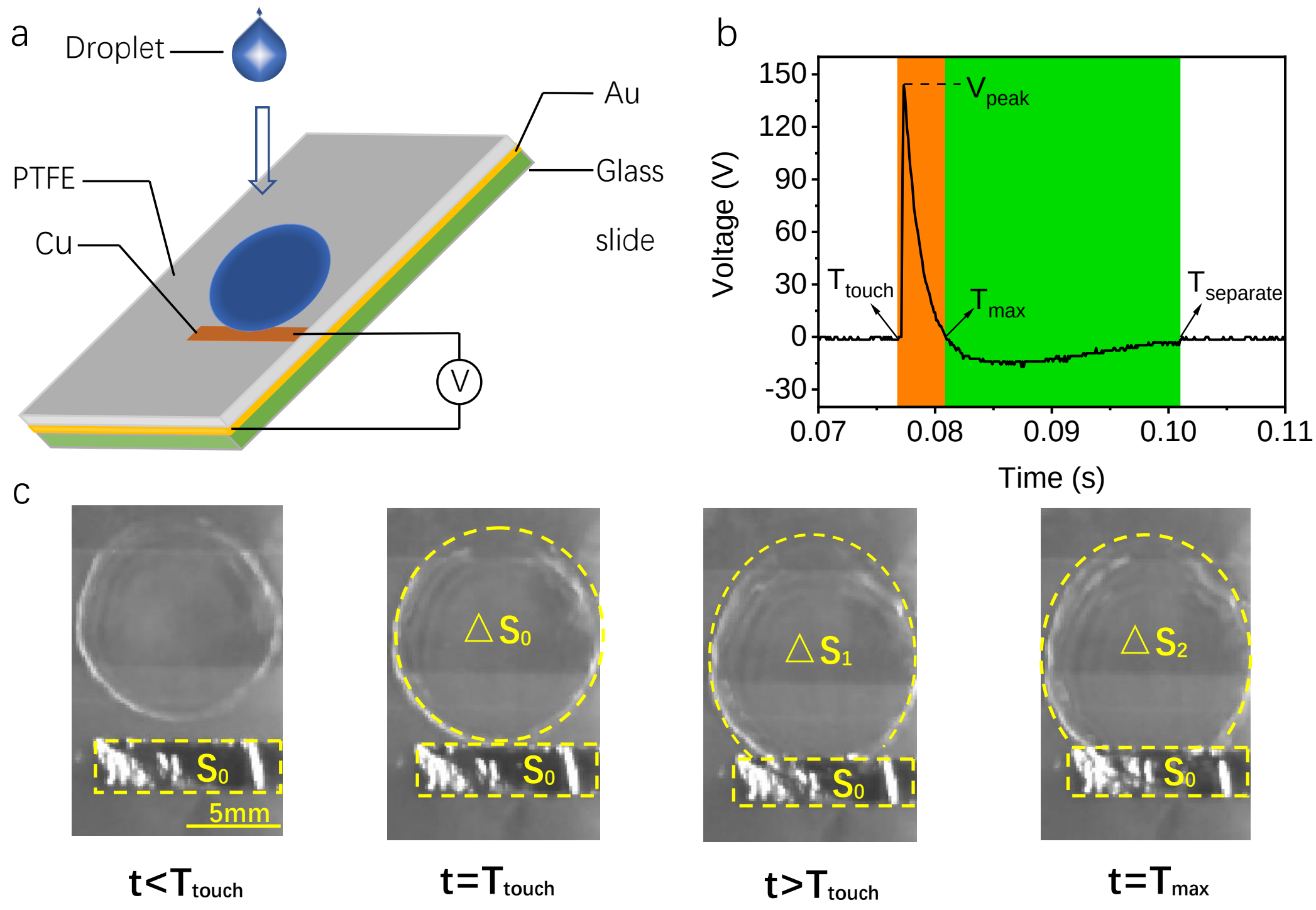

**Figure 1. Voltage pulse generated by dropping a droplet onto the device. (a)** Schematic illustration of the droplet-based generator. **(b)** Voltage pulse generated by dropping one droplet. $T_{touch}$, $T_{max}$, $T_{separate}$ denote the time one droplet just touches the top electrode and the voltage pulse appears, the time droplet spreads to maximum area and the time droplet separated from the electrode respectively. Orange region: time interval of droplet spreading. Green region: time interval of droplet shrinking. **(c)** Observation of the dynamics of the droplet. Here $S_0$, $\Delta S$ denote the area of the top electrode and the droplet area subtracted by the area covered with the top electrode respectively.

To reveal the relationship between the output signal and the dynamical change of the droplet area, we took a close-up view through a high-speed camera, it was discovered that the area change rate of the spreading droplet $dS/dt$ shows a similar-shaped curve compared with the voltage pulse, which suggested there is a close relationship between induced voltage and *dS/dt*.

The *dS/dt* and output voltage peak ($V_{peak}$) also show similar dependence on factors including the height (*H*) and frequency of the released droplet, as well as tilt angle (α) of the (inset of Fig. 2b). As shown in Fig. 2b, the droplet released from a higher height causes a larger maximum of *dS/dt* as a result of the larger velocity when touching the electrode. Meanwhile, the induced voltage peak increases with the increase of the height, showing similar trend.

When the droplets are released at a faster frequency, *dS/dt* has a larger maximum. This is due to the lack of time for the water film to leave the surface completely before the next droplet falls, and the residue of the water film acting as a lubricant, leading to a faster change rate of water area. As the frequency increases, there is a coincident increase compared with the maximum of *dS/dt*.

When the tilt angle of the device increases, the voltage peak has a variation in accordance with the maximum of *dS/dt*, and both of them reach the maximum at around 45°. Under the observation by a high-speed camera, it's easy to see that the shape and area of the spreading droplet change significantly at different tilt angle (Supplementary Fig. 1).

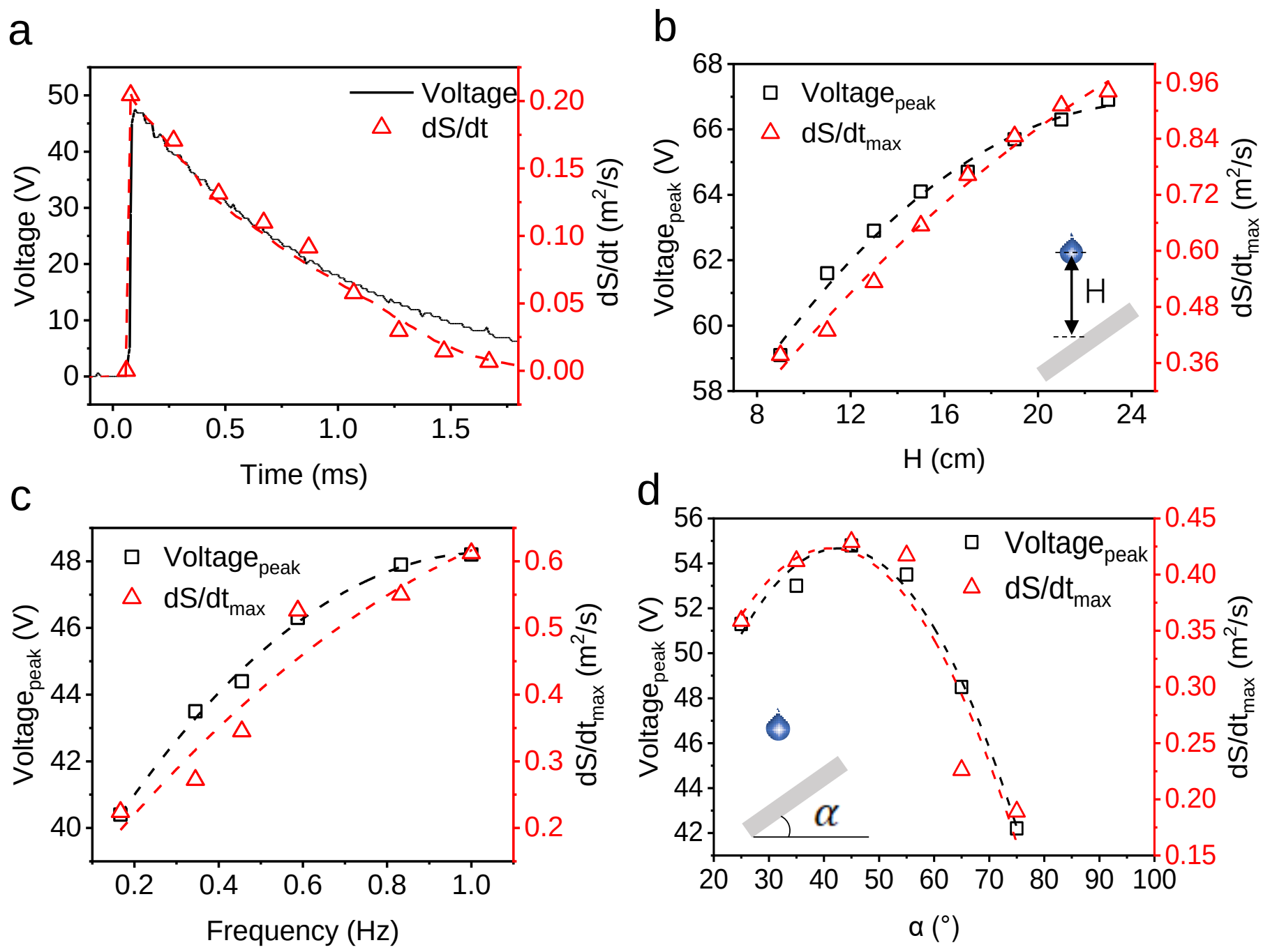

**Figure 2. The area changing rate *dS/dt* and corresponding voltage signal. (a)** Evolution of *dS/dt* from the droplet contacts the top electrode to it reaches the maximum area (*dS/dt* = 0*)*. The output voltage shows similar shape curve. **(b)** The dependence of $Voltage_{peak}$ and $dS/dt_{max}$ on the height (*H*) at which the droplets were released. **(c)** The dependence of $Voltage_{peak}$ and $dS/dt_{max}$ on the frequency at which the droplets were released. **(d)** The dependence of $Voltage_{peak}$ and $dS/dt_{max}$ on the tilt angle (α) of the device.

## Mechanism

The DEG could be considered as a parallel-plate capacitor, in which the parallel plates are the top electrode and the bottom electrode, and the dielectric layer is the PTFE film. The stable negative charges[21-22] stored on the surface of the PTFE film cause an electric field $E_{PTFE}$ between the two plates, and $E_{PTFE}$, which is equal to $\sigma_{PTFE}/(2\varepsilon)$ according to Gauss' law, where $\sigma_{PTFE}$ is surface charge density of the PTFE film and $\varepsilon$ is permittivity of PTFE. The presence of $E_{PTFE}$ drives electron migrates between the two plates. During the migration progress, the electric field $E_{plate}$ generated by charges in the two plates is increasing, and $E_{plate}$ is equal to $\sigma_{plate}/\varepsilon$, where $\sigma_{plate}$ is surface charge density of the plates. Once $E_{plate}$ increases to equal to $E_{PTFE}$, electrons stop directional migrating as a result of electric equilibrium (Fig. 3a). In this electric equilibrium, surface charge density $\sigma_{plate}$ and surface charge density $\sigma_{PTFE}$ is required to satisfy following equations:

$$\frac{\sigma_{PTFE}}{2\varepsilon} = \frac{\sigma_{plate}}{\varepsilon} \tag{1}$$

$$2\sigma_{plate} = \sigma_{PTFE} = \frac{2Q_0}{S_0} \tag{2}$$

Where $Q_0$ is the number of charges of the plates in this electric equilibrium, and $S_0$ is the area of the top electrode.

Once one droplet drops onto the PTFE surface and contacts the top electrode, the droplet will spread around the top electrode, which means the area of water around the top plate is increasing. The droplets are tap water that contains ions with electrical conductivity. The conductivity of the droplets suggests that it's reasonable to treat the increasing area of the droplet as an equivalent extra part $\Delta S$ of the capacitor top plate. Therefore, the total area of the top plate *S* now equals to the top electrode area $S_0$ pluses the equivalent extra area $\Delta S$ (Fig. 1c). Then the surface charge density $\sigma_{plate}$ can be expressed as:

$$2\sigma_{plate} = \frac{2Q_0}{\varepsilon\ (S_0+\Delta S)} < \sigma_{PTFE} \tag{3}$$

As shown in Fig.3b, the equation (3) indicates that $2\sigma_{plate} < \sigma_{PTFE}$ because the extra $\Delta S$ appears in the denominator. The electric equilibrium has been broken. In other words, the extra area $\Delta S$ makes the electric field $E_{plate}$ weaker and not compete with the electric field $E_{PTFE}$, as a result of a sparser electric charge density of the plates (Fig. 3b). To recover the electric equilibrium, the only way is to increase *Q*. So electrons begin migrating from the top plate to the bottom plate again until $2\sigma_{plate} = \sigma_{PTFE}$ (see Fig.3c). The increment $\Delta Q$ should follow the following equations:

$$2\sigma_{plate} = \frac{2(Q_0+\Delta Q)}{\varepsilon(S_0+\Delta S)} = \sigma_{PTFE} \tag{4}$$

$$\frac{\Delta Q}{\Delta S} = \frac{\varepsilon\sigma_{PTFE}}{2} \tag{5}$$

According to the definition of current, we can calculate current *I* and voltage *U* by *dS/dt*:

$$I = \frac{dQ}{dt} = \frac{\varepsilon\sigma_{PTFE}}{2}\frac{dS}{dt} \tag{6}$$

$$U = IR = \frac{R\varepsilon\sigma_{PTFE}}{2}\frac{dS}{dt} \quad (7)$$

Here, $R$ is the internal resistance. Equation (6) and equation (7) directly indicate the relationship between electricity generated by droplets and the area change rate of the spreading droplet: $I$ and $U$ are directly proportion to $dS/dt$ and the proportionality coefficient is mainly determined by surface charge density of the electret material.

When the droplet was just in contact with the top electrode, $\Delta t$ is close to 0 and $\Delta S$ suddenly goes from 0 to the spreading droplet area, so the voltage peak is theoretically infinite. However, on the micro scale, a barrier existing at the interface between the droplet and the top electrode and the resistance at this interface and inside the droplet to charge transfer limit the voltage peak. To achieve a higher voltage peak, appropriate optimization is crucially required, especially at the interface between the droplet and the electrode.

The droplet cannot always keep spreading, and in quick succession, the droplet begins shrinking after the droplet spreads to the maximum area. During the shrinking progress, the extra area around the top electrode $\Delta S$ is decreasing. This gives rise to a reversed progress: the decreasing $\Delta S$ causes the increment of surface density of the two plates. Therefore, $\sigma_{PTFE}$ is increasing and breaks the electric equilibrium again. To recover the electric equilibrium, electrons will migrate from the bottom plate to the top plate in a reversed direction. These reversed electrons migration means a back flow of current and a negative voltage pulse that has mentioned in Fig.1b.

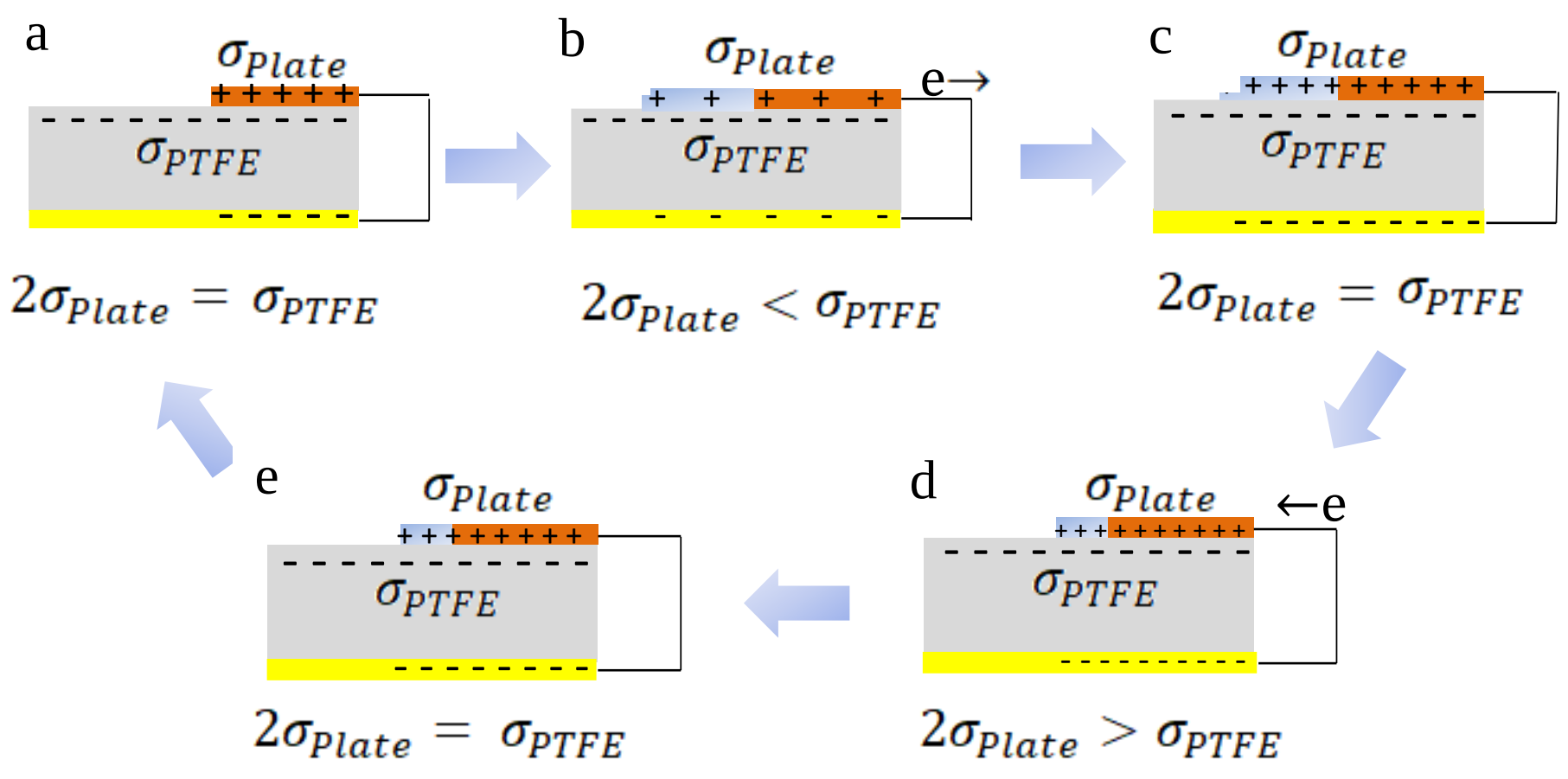

**Figure 3. Schematic illustration of the mechanism. (a)** The equilibrium state when no droplet on the surface, $2\sigma_{plate} = \sigma_{PTFE}$. **(b)** State when a droplet in contact with the top electrode is spreading, $2\sigma_{plate} < \sigma_{PTFE}$, change of $\Delta S$ breaks the equilibrium and electrons migrate from the top plate to the bottom plate. **(c)** The electric equilibrium is recovered until $2\sigma_{plate} = \sigma_{PTFE}$. **(d)** The shrinking droplet causes $\Delta S$ decrease, $2\sigma_{plate} < \sigma_{PTFE}$, electrons are driven to migrate in a reversed direction.

**The drop location.**

Having understood the underlying mechanism, we know that *dS/dt* is the key factor that dominates the induced voltage. Therefore, the location of the droplet touching the PTFE film surface should be seriously considered, because it affects $\Delta S$ significantly. The shape of the droplet changes with the location and can be classified into complete shape, incomplete shape and dew shape respectively, corresponding to inset in Fig. 4a, inset in Fig. 4b and inset in Supplementary Fig. 2. The complete shape has the largest *dS/dt* that is about twice as much as the area of the incomplete shape, and the dew shape has the smallest *dS/dt*. According to our model, the larger *dS/dt* generates a higher voltage peak. The measurements are in good agreement with the proposed model: the droplet of complete shape induces a highest voltage peak (Fig. 4a) that is just about twice as the voltage peak generated by the droplet of incomplete shape (Fig, 4b), and the droplet of dew shape induces the smallest voltage peak (Supplementary Fig, 2). To reach a higher voltage peak, we should make sure the droplet drops on the surface and contacts the electrode in a complete shape.

**Surface charge density of PTFE film.**

According to equation (6) and equation (7), surface charge density $\sigma_{PTFE}$ decides the generated voltage at the same *dS/dt.* Voltage will be increased if $\sigma_{PTFE}$ is larger. To verify the role of $\sigma_{PTFE}$ in the proposed model, we treated the device using plasma that can increases $\sigma_{PTFE}$, which leaded a larger $\sigma_{PTFE}$. As shown in Fig. 4c, the peak of generated voltage pulse is about twice as big as the one without plasma treatment (Fig. 4d), which is in agreement with our model. Therefore, to further develop the capability of electricity generation, injecting ions to the electret material is a promising mean if the injected ions can exist steadily on the surface.

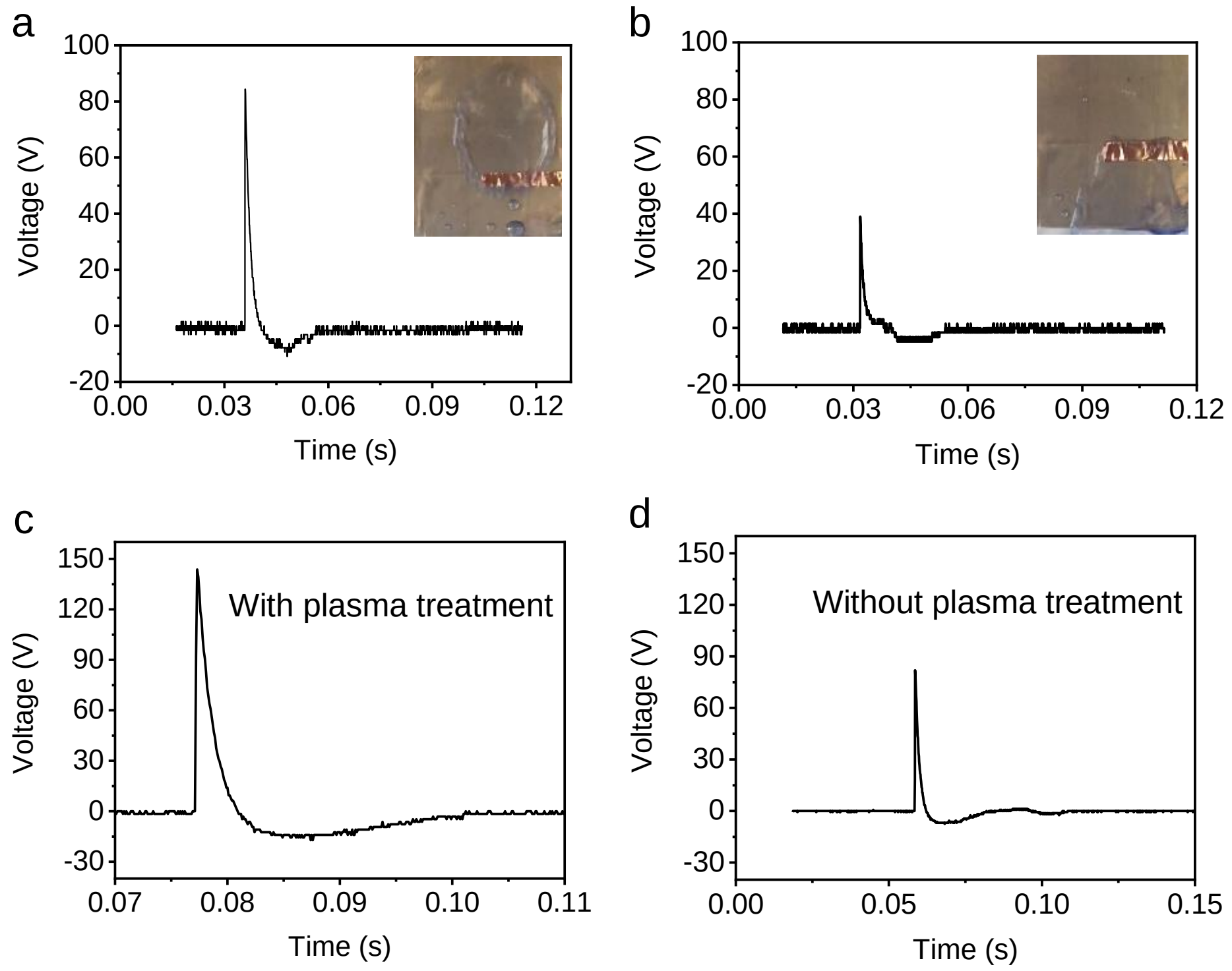

**Figure 4. Factors that affect the output voltage. (a, b)** Voltage signals for droplets released onto PTFE surface above (a) and below (b) the top electrode, respectively. To clearly show the shape of the droplet, we used diluted blue ink to take place of tap water. **(c)** The voltage generated by a device pretreated by plasma. **(d)** The voltage generated by a device without plasma treatment.

## Conclusions

The dynamic model of surface charge density based on a series of comprehensive experiments sheds light on the underlying working mechanism of the DEGs. The DEG serve as a parallel-plate capacitor and the droplet contacting the top electrode can be equivalently regarded as an additional part of the top plate. When the droplet area $\Delta S$ changes, the electrical equilibrium is broken simultaneously because of changed surface charge density of the plates. To match up stable surface charge density of the electret material, electrons are driven to migrate between the two conductive plates until the electric equilibrium is recovered. The voltage and current are derived to be directly proportionally to the area change rate *dS/dt* of the droplet, and surface charge density of the electret material determines the proportionality coefficient. The change rate of the droplet area *dS/dt* is a mathematical manifestation of the droplet moving boundary, and the model of surface charge density could be considered as a ramification of the moving boundary model proposed previously[1-2]. Under the guidance of the insight, the DEGs can be further developed to harvest the mechanical energy of droplets more efficiently.

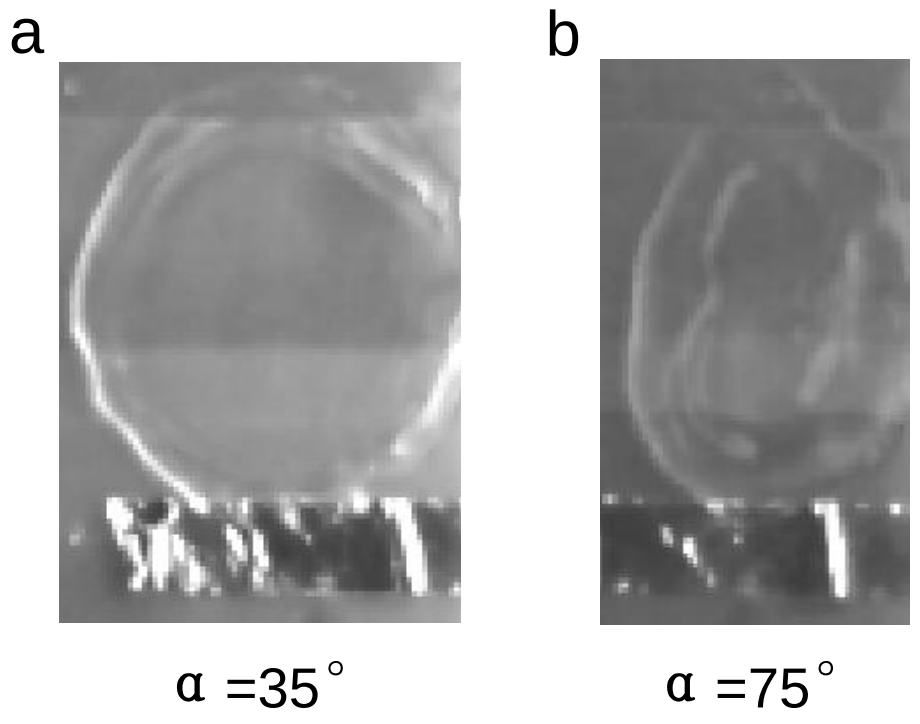

**Supplementary Figure 1. The shape of the spreading droplet at different inclinations. (a)** The shape and area of the spreading droplet at 35°. **(b)** The shape and area of the spreading droplet at 75°.

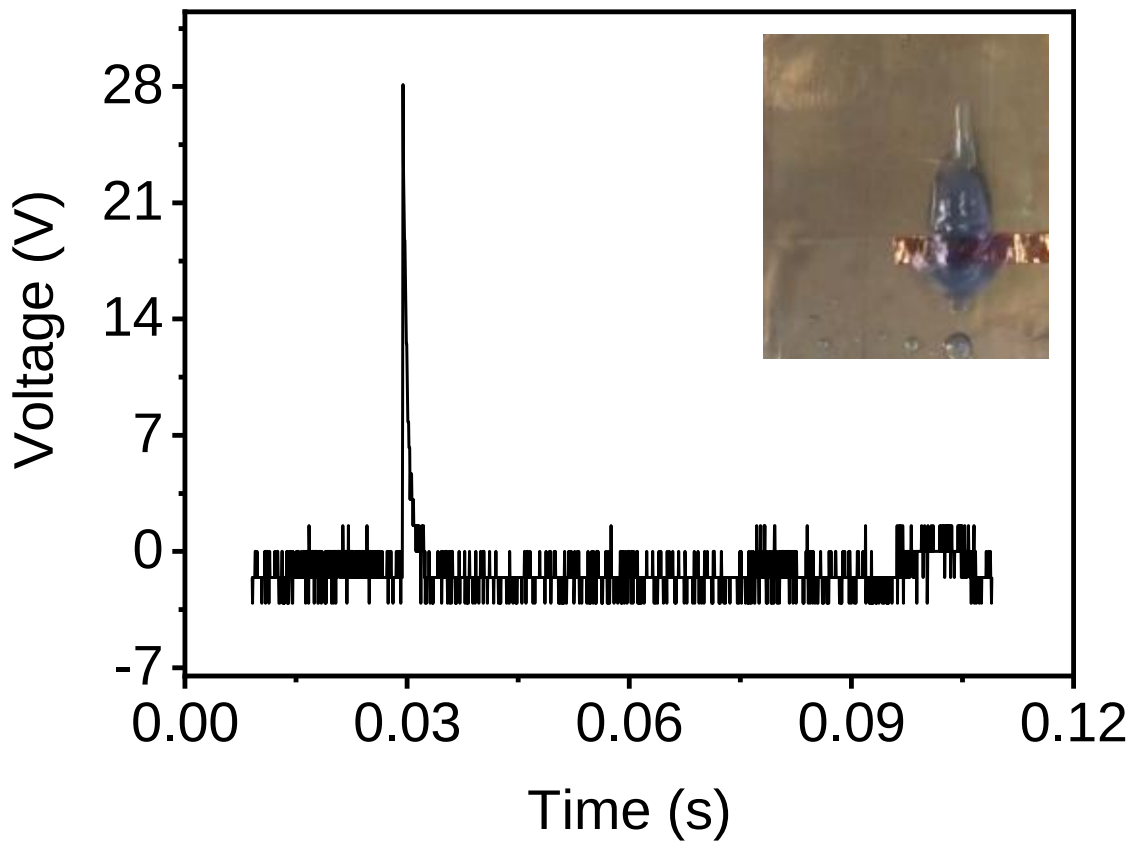

**Supplementary Figure 2. Voltage signal for sliding droplets.** Voltage signals for droplets gliding to the top electrode. To clearly show the shape of the droplet, we used diluted blue ink to take place of tap water.